\documentclass[preprint,aps,draft]{revtex4}
\usepackage{graphicx}
\usepackage{amsfonts}
\usepackage{amssymb}%
\newcommand{\f}{\begin{equation}}
\newcommand{\ff}{\end{equation}}
\newcommand{\fa}{\begin{eqnarray}}
\newcommand{\ffa}{\end{eqnarray}}
\begin{document}

\preprint{}

\title{Modified (A)dS Schwarzschild black holes in Rainbow spacetime}% Force line breaks with \\

\author{Huarun Li}
 \email{alloys91@sohu.com}
\author{Yi Ling}
 \email{yling@ncu.edu.cn}
\author{Xin Han}%
 \email{bifrostx@gmail.com}
\affiliation{ Center for  Gravity and Relativistic Astrophysics,
Department of Physics, Nanchang University, Nanchang 330031, China
}%

\begin{abstract}
A modified (Anti-)de Sitter Schwarzschild black hole solution is
presented in the framework of rainbow gravity with a cosmological
constant. Its thermodynamical properties are investigated. In
general the temperature of modified black holes is dependent on the
energy of probes which take the measurement. However, a notion of
intrinsic temperature can be introduced by identifying these probes
with radiation particles emitted from black holes. It is interesting
to find that the Hawking temperature of this sort of black holes can
be reproduced by employing the extended uncertainty principle and
modified dispersion relations to the ordinary (A)dS Schwarzschild
black holes.

\end{abstract}

\maketitle

\section{Introduction}
\baselineskip=20pt

Doubly Special Relativity (DSR) has gradually been viewed as an
effective theory to describe physics phenomena at extremely high
energy level when the semi-classical effects of quantum gravity is
taken into
account\cite{AmelinoCamelia00mn,Magueijo01cr,Magueijo02am,AmelinoCamelia03ex,Girelli04md,
AmelinoCamelia08qg,Glikman04qa,Smolin05cz}. In this framework the
Lorentz symmetry may be preserved or deformed. One of its key
ingredients is that the usual energy-momentum or dispersion relation
may be modified with corrections in the order of Planck length $l_p=
\sqrt{8 \pi G}\sim 1/M_p$
\begin{equation}
E^2f^2(l_pE)-P^2g^2(l_pE)=m^2,\label{mdr1}
\end{equation}
where $f(E)$ and $g(E)$ are two general functions of energy with a
constraint that they approach to unit as the energy of particles is
much less than the Planck energy $M_p$. Modified dispersion
relations (MDR) and its implications to physics have been greatly
investigated in recent years\cite{AmelinoCamelia00mn,Liberati07vx}.
They may be responsible for threshold anomalies of ultra high energy
cosmic rays and Gamma ray
burst\cite{Colladay98fq,Coleman98ti,Amelino00zs,Jacobson01tu,Myers03fd,Jacobson03bn,Abbasi07sv},
contribute corrections to the entropy of black holes
\cite{AmelinoCamelia:2005ik,Ling05bq}, provide alternatives to
inflationary cosmology\cite{Moffat92ud,Albrecht98ir}, and disclose
the nature of dark energy\cite{MersiniHoughton:2001su}.

Recently DSR has been generalized to incorporate the effects of
gravity, leading to a deformed formalism of general relativity,
which is always named as rainbow gravity because the metric of the
background detected by any probe is not fixed, but depends on the
energy of the probe\cite{Magueijo:2002xx}. Specifically, given a
modified dispersion relation as in Eq.(\ref{mdr1}), the usual flat
metric is replaced by a one-parameter family of rainbow metric
\begin{equation}
ds^2=-{1\over f^2(E)}dt^2+{1\over
g^2(E)}dx^2,\label{rfs}\end{equation} such that the contraction
between momentum and infinitesimal displacement
\begin{equation} dx^{\mu}p_{\mu}=dtE+dx^ip_i,\end{equation}  be an invariant
quantity. This strategy overcomes the difficulty of defining the
position space conjugate to momentum space arising in $DSR$ where
the Lorentz symmetry is accomplished by nonlinear transformations in
momentum space\footnote{For recent progress and discussion on this
issue, we may refer to \cite{Hossenfelder06rr}.}. When the curvature
of spacetime is taken into account, a deformed equivalence principle
of general relativity is proposed, requiring that the free falling
observers who make measurements with energy $E$ will observe the
same laws of physics as in doubly special relativity. As a
consequence, in rainbow gravity the original Einstein field equation
is replaced by a one-parameter family of modified equations,
\begin{equation} G_{\mu\nu}(E)=8\pi
G(E)T_{\mu\nu}(E)+g_{\mu\nu}(E)\Lambda (E),\label{MEE}\end{equation}
 where Newton's constant as well as
the cosmological constant is conjectured to be energy dependent as
one expects from the viewpoint of renormalization group
theory\cite{Reuter96cp,Girelli06sc}.

This formalism has received a lot of attention and some relevant
work can be found in
Ref.\cite{Galan04st,Galan05ju,Hackett05mb,Aloisio05qt,Ling05bp,Galan06by,Ling06az,
Ling:2006ba,Girelli06fw,Leiva08fd,Grillo08zz,Peng07nj}. In
particular in \cite{Ling05bp} one of our authors with other
collaborators studied the thermodynamics of modified Schwarzschild
black holes appearing in rainbow gravity. Although the temperature
of such kind of black holes is also energy dependent, a notion of
intrinsic temperature was introduced by identifying probes with
those photons surrounding the vicinity of black hole horizon.

In this short report we intend to extend this strategy to rainbow
gravity with a cosmological constant. We firstly show in section two
that there exists a modified (A)dS Schwarzschild black hole solution
to the modified vacuum field equation, then turn to investigate the
thermodynamics of this sort of black holes in section three. Given a
specific modified dispersion relation, the intrinsic temperature can
be obtained. Comparing it with the one obtained in Ref.\cite{HLL},
where the impact of modified dispersion relation on ordinary
(A)dS-Schwarzschild black holes is investigated, we find that they
are exactly the same such that the proposal of deformed equivalence
principle is testified in this context.

\section{The modified (A)dS Schwarzschild Solution}
In this section we demonstrate a spherically symmetric solution to
the modified field equations (\ref{MEE}) with non-zero
cosmological constant. In the absence of matter, it reduces to
$G_{\mu\nu}(E)=\Lambda(E) g_{\mu\nu}$, where $\Lambda(E) =
\pm3/L^2(E)$ is an energy-dependent cosmological constant, where
$``+"$ sign and $``-"$ sign correspond to dS and AdS cases
respectively. For explicitness let us focus on the AdS black holes
in following discussions but all derivations can be applied to the
de-Sitter case directly. Given a general modified dispersion
relation as shown in (\ref{mdr1}), we may write the most general
form for a spherically symmetric metric in energy independent
coordinates as
\begin{equation}
\ ds^2 = - \frac{e^{2a(r)}}{f^2(E)} d\tau^2 +
\frac{e^{2b(r)}}{g^2(E)} dr^2 + \frac{r^2}{g^2(E)} (d \theta ^2 +
\sin ^2 \theta d \phi ^2 ),\label{eq5}
\end{equation}
where $a(r)$and $b(r)$ are two functions we intend to solve for.
They may be energy-dependent.

Now it is straightforward to obtain the non-vanishing components
of the connection as follows
\begin{equation}
\begin{array}{llll}
\Gamma ^0_{10} = \Gamma ^0_{01} = a', &\Gamma ^1_{00} =
g^2(E)f^{-2}(E) e^{2(a-b)} a', &\Gamma ^1_{11} = b', \nonumber\\
\Gamma ^1_{22} = -re^{-2b}, &\Gamma ^1_{33} = -r \sin ^2 \theta e^{-2b}, &\Gamma ^2_{12}=\Gamma^2_{21}=r^{-1},\nonumber\\
\Gamma ^3_{13}=\Gamma ^3_{31}=r^{-1},& \Gamma ^2_{33}=- \sin \theta
\cos \theta, &\Gamma^3_{23}=\Gamma^3_{32}= \cot \theta, \end{array}
\end{equation}
where a prime sign denotes the derivative with respect to the
radius $r$. Thus the non-vanishing Ricci tensor components are
\begin{equation}
\begin{array}{llll}
R_{00} &=& \frac{g^2(E)}{f^2(E)}e^{2(a-b)}[a''-a'b'+
a'^2+2r^{-1}a'],\\
R_{11} &=& -[ a''-a'b'+a'^2-2r^{-1}b' ],\\
R_{22} &=& -e^{-2b}[1+r(a'-b')]+1,\\
R_{33} &=& \{-e^{-2b}[1+r(a'-b')]+1\} \sin ^2 \theta,
\end{array}
\end{equation}
and the Ricci scalar is
\begin{equation}
R=2g^2(E)r^{-2}+2g^2(E)e^{-2b}[(-a''+a'b'-a'^2)-r^{-2}-2r^{-1}(a'-b')].
\end{equation}

Substituting all these terms into the AdS vacuum field equations
gives rise to
\begin{equation}
-g^2(E)e^{-2b}(2r^{-1}b'-r^{-2})-g^2(E)r^{-2}=\frac{3}{L^2(E)},\label{eq9}
\end{equation}
\begin{equation}
g^2(E)e^{-2b}(2r^{-1}a'+r^{-2})-g^2(E)r^{-2}=\frac{3}{L^2(E)},\label{eq10}
\end{equation}
\begin{equation}
g^2(E)e^{-2b}[a''-a'b'+a'^2+r^{-1}(a'-b')]=\frac{3}{L^2(E)}.\label{eq11}
\end{equation}

From (\ref{eq9}) and (\ref{eq10}) we can obtain a relation between
$a$ and $b$ as
\begin{equation}
a=-b+\alpha,
\end{equation}
with $\alpha$ being a constant. Furthermore we can obtain
$e^{-2b}$ from (\ref{eq9}) and (\ref{eq10}) as
\begin{equation}
e^{-2b}=1+cr^{-1}+g^{-2}(E)L^{-2}(E)r^2,
\end{equation}
where $c$ is an integral constant.

As a result, the rainbow metric in eq.(\ref{eq5}) becomes
\begin{equation}
ds^2=-\frac{N^2e^{2\alpha}}{f^2(E)}d\tau^2+\frac{1}{g^2(E)N^2}dr^2+
\frac{r^2}{g^2(E)}(d \theta ^2 +\sin ^2 \theta d \phi ^2
),\label{eq14}
\end{equation}
where $N^2=1+cr^{-1}+g^{-2}(E)L^{-2}(E)r^2$.

As usual we can define a new coordinate by $t=e^{\alpha}\tau$ such
that the metric becomes

\begin{equation}
ds^2=-\frac{N^2}{f^2(E)}dt^2+\frac{1}{g^2(E)N^2}dr^2 +
\frac{r^2}{g^2(E)}(d \theta ^2 + \sin ^2 \theta d \phi ^2
).\label{eq15}
\end{equation}

Now the remaining task is to determine the constant $c$. Firstly
we notice that as the cosmological constant vanishes, this
solution should recover the modified Schwarzschild solution which
has been obtained in \cite{Magueijo:2002xx}. This directly leads
to $c=-2G(0)M$. As a matter of fact we point out that this fixing
implies that we have set the energy dependent Newton's constant as
$G(E)=g^{-1}(E)G(0)$. When the cosmological constant is not zero,
we argue that it is appropriate to set $\Lambda
(E)=g^{2}(E)\Lambda(0)$ based on the following considerations.
Given a rainbow metric as Eq.(\ref{eq5}), we notice that in
general the volume of a spatial region with fixed size $R$ is
energy dependent as $V\sim g^{-3}(E)V_0$. It is reasonable to
require that the total vacuum energy in this region is independent
of the energy of free-falling probes, namely
\begin{equation} E_{vac}\sim \rho(E)V(E)\sim \frac{\Lambda(E)}{8\pi
G(E)}g^{-3}(E)V_0\sim Const.\end{equation}

Once Newton's constant is set as $G(E)=g^{-1}(E)G(0)$, we find the
condition above is satisfied if
\begin{equation}
\Lambda(E)=g^{2}(E)\Lambda(0)=\pm\frac{3g^{2}(E)}{L^2(0)}.\nonumber
\end{equation}

Finally the rainbow metric for modified AdS Schwarzschild black
holes takes the form

\begin{equation}
ds^2=-\frac{N^2}{f^2(E)}dt^2+\frac{1}{g^2(E)N^2}dr^2+\frac{r^2}{g^2(E)}(d\theta
^2+\sin ^2 \theta d \phi ^2),\label{eq18}
\end{equation}
where $N^2=1-2G(0)Mr^{-1}+L^{-2}(0)r^2$.

It is manifest that at low energy limit, namely
$E/M_{p}\rightarrow 0$, this solution becomes the conventional
anti-de-Sitter Schwarzschild vacuum solution.
\begin{equation}
ds^2=-N^2dt^2+\frac{1}{N^2}dr^2+r^2(d\theta ^2+\sin ^2 \theta d \phi
^2).\label{eq19}
\end{equation}

In parallel the modified dS Schwarzschild black hole solution to the
modified vacuum field equation can be easily obtained by considering
$L^2(0)\longrightarrow -L^2(0)$ in (\ref{eq18}) and it can return to
the ordinary dS Schwarzschild vacuum solution at the low energy
limit.

In these solutions, the position of horizon is determined by
\begin{equation} 2G(0)M=r_+\left(1\pm\frac{r^2_+}{L^2(0)}\right)
\label{eq20},
\end{equation}
where $``+"$ sign and $``-"$ sign correspond to AdS and dS black
holes respectively. Thus, in general the area of the horizon is
energy dependent as $A=(4\pi r^2_+)/g^2(E)$.

\section{Thermodynamics with the modified (A)dS Schwarzschild black holes}

Now we turn to investigate the thermodynamics of the modified (A)dS
Schwarzschild black holes. As pointed out in \cite{Ling05bp}, in the
framework of rainbow gravity, the temperature of the modified (A)dS
Schwarzschild black holes (\ref{eq18}) can be identified with the
surface gravity $\kappa$ on the horizon, namely $ T=\kappa/(2\pi)$,
where $\kappa$ is related to the metric by

\begin{equation}
\kappa=-\frac{1}{2}\lim_{r\longrightarrow r_+}
{\sqrt{\frac{-g^{11}}{g^{00}}}}\frac{(g^{00})'}{g^{00}}\label{eq21}
.
\end{equation}

Now it is straightforward to obtain the surface gravity for modified
Schwarzschild (A)dS black holes as

\begin{equation}
\kappa=\frac{g(E)}{2f(E)}\left(\frac{2GM}{r^2_+}\pm\frac{2r_+}{L^2(0)}\right)
=\frac{g(E)}{2f(E)}\left(\frac{1}{r_+}\pm\frac{3r_+}{L^2(0)}\right)\label{eq22},
\end{equation}
where $``+"$ sign and $``-"$ sign correspond to AdS and dS black
holes respectively. Thus, the temperature reads as

\begin{equation}
T_{(A)dS}=\frac{\kappa}{2\pi}=\frac{g(E)}{4\pi
f(E)}\left(\frac{1}{r_+}\pm\frac{3r_+}{L^2(0)}\right)\equiv\frac{g(E)}{f(E)}T_{0(A)dS},
 \label{eq23}\end{equation}
where $T_{0(A)dS}$ is understood as the temperature for the ordinary
(A)dS black holes, i.e.
\begin{equation}
 T_{0(A)dS}=\frac{1}{4\pi}\left(\frac{1}{r_+}\pm\frac{3r_+}{L^2(0)}\right)\label{eq24},
\end{equation}
Equation $(\ref{eq23})$ indicates that the temperature of the
modified black holes is different for probes with different energy.
Now following the strategy in \cite{Ling05bp} we propose to define
an intrinsic temperature for large modified black holes by
identifying probes with radiation particles in the vicinity of black
hole horizon. Suppose we make the measurement with the use of
radiation photons with average energy $E=<E>$, then we expect that
the temperature of black holes can be identified with the energy of
photons emitted from black hole
\cite{Ling05bp,Galan06by,Adler:2001vs,Chen:2002tu,Scardigli06eb},
namely $T = E$. This identification provides a scheme to define an
intrinsic temperature for modified black holes. For explicitness, we
consider a specific modified dispersion with $ f^2=1-(l_pE)^2 $ and
$g^2=1$. In this case plugging $T=E$ into equation $(\ref{eq23})$ we
have

\begin{equation}
l^2_pT^4-T^2+T^2_0=0\label{eq25}.
\end{equation}

As a result, the intrinsic temperature of modified Schwarzschild
(A)dS black holes is obtained as

\begin{equation}
T_{(A)dS}=\left[\frac{M^2_p}{2}\left(1-\sqrt{1-\frac{4T^2_0}{M^2_p}}\right)\right]^{\frac{1}{2}}
\label{eq26}.
\end{equation}

Obviously for large black holes with $2T_0\ll M_p$, the temperature
goes back to the ordinary form $T\sim T_0$, while for small black
holes with extremely high temperature, we find it reaches to a
maximal value $T\sim M_p/\sqrt{2}$ as $T_0\sim M_p/2$.
Correspondingly the radius of the black hole horizon is also bounded
from below with a constraint as
\begin{equation}
\frac{1}{4\pi}(\frac{1}{r_+}\pm\frac{3r_+}{L^2(0)}) \leq
M_p/2,\label{eq27}
\end{equation}
implying  $r_+ > {\frac{l_p}{2\pi}}$.

The existence of a minimum radius implies that the black hole might
stop radiating at the late moment of evaporation. As a matter of
fact, this can be seen from the change of the heat capacity of (A)dS
Schwarzschild black holes, which is defined as
\begin{equation}
C_{(A)dS}=\frac{dM}{dT}.\label{eq28}
\end{equation}

From (\ref{eq20}) and (\ref{eq26}), we obtain it as
\begin{equation}
C_{(A)dS}=-\frac{2\pi\sqrt{1-4l^2_pT^2_0}\left(1\pm\frac{3r^2_+}{L^2(0)}\right)}
{G(0)\sqrt{1-l^2_pT^2}\left(\frac{1}{r^2_+}\mp\frac{3}{L^2(0)}\right)},\label{eq29}
\end{equation}
where the upper sign and lower sign correspond to AdS and dS black
holes respectively.

Thus we find that the heat capacity becomes vanishing when the black
hole temperature reaches its maximum value, i.e.
$T=\frac{M_p}{\sqrt2}$. This result implies that there might exist a
ground state of modified black holes with non-zero mass such that it
may provides a mechanism to take black hole remnants as a candidate
for cold dark matter due to their weakly interacting features
\cite{MacGibbon:1987my,Barrow:1992hq,Carr:1994ar}.

In the remainder of this section we point out that the intrinsic
temperature for modified black holes we obtained above, namely
equation $(\ref{eq23})$, is consistent with the one for ordinary
(A)dS black holes which are supposed to be surrounded by radiation
particles endowed with modified dispersion relations. The
derivation there can be summarized as follows. Given a modified
dispersion relation as $(\ref{mdr1})$, we apply it to the photons
emitted from $\it ordinary$ (A)dS black holes and identify the
characteristic temperature of this black hole with the averaged
energy $E$, we may have \begin{equation}
T^{EUP}_{(A)dS}=\frac{g(E)}{ f(E)}P. \label{temEUP}\end{equation}

Moreover, we apply an extension of the ordinary uncertainty relation
to photons in the vicinity of black hole
horizon\cite{Bolen,Park07az}, i.e., \f P\sim \delta P\sim
{1\over\delta x}\left( 1+\beta^2\frac{\delta x^2}{L^2}\right)\sim
{1\over 4\pi r_+}\left( 1+\beta^2\frac{r_+^2}{L^2}\right),\ff where
a ``calibration factor'' $4\pi$ as well as a dimensionless parameter
$\beta$ is introduced. Plugging this relation into (\ref{temEUP})
and properly setting the value of $\beta$ we easily find the
temperature of (A)dS Schwarzschild black holes has the same form as
(\ref{eq23}). For details we refer to Ref.\cite{HLL}. Here we stress
that this consistency is universal and independent of the specific
form of dispersion relations. Therefore, our results obtained here
in the context of modified (A)dS black holes may be viewed as a
powerful support for the proposal of deformed equivalence principle
in \cite{Magueijo:2002xx}.

\section{Summary}

In this paper we investigated (A)dS Schwarzschild black holes in the
framework of rainbow gravity with a cosmology constant. We firstly
derived a modified (A)dS Schwarzschild black hole solution to the
modified vacuum field equation, then studied their thermodynamical
properties. In general the temperature of the modified black holes
can be obtained by calculating the surface gravity on the horizon.
It turns out that it depends on the energy of probes but goes back
to the standard result for ordinary (A)dS Schwarzschild black holes
as the energy is much less than the Planck energy. When identifying
probes as those photons surrounding the horizon of black holes, we
also introduced a notion of intrinsic temperature for these modified
black holes. In particular, we considered the thermodynamics of a
special sort of rainbow black holes by specifying the functions
$f(E)$ and $g(E)$. Through our investigations the following two
important conclusions may be drawn. First our results for modified
(A)dS black holes coincide with those for ordinary (A)dS
Schwarzschild black holes, where the temperature is obtained by
employing modified dispersion relation and extended uncertainty
principle to photons directly. This equivalence testifies the
proposal of deformed equivalent principle, namely freely falling
observers in rainbow spacetime have the same physics laws as those
in doubly special relativity. Second, The modification of  (A)dS
Schwarzschild black holes may provide a reasonable scenario for
understanding the late fate of black hole evaporation because in
this formalism the evaporation may stop as the heat capacity
vanishes. At that moment the temperature of black holes is extremely
high, but not divergent as occurring in the ordinary picture of
Hawking radiation. Thus the black hole radiation would stop with a
remnant which may be viewed as a candidate for mysterious dark
matter.

Finally it is worthwhile to point out that our derivations are
mainly in a heuristic manner, particularly in the part of the
thermodynamics of modified $AdS$ black holes. Firstly we have
assumed that the temperature of modified black holes can still be
related to the surface gravity as usual, and the reasons that we can
do so have been given in\cite{Ling05bp}. To strictly derive or prove
this relation one need a quantum field theory on rainbow spacetime,
which as far as we know is still absent, though some relevant work
can be found, for instance in Ref.\cite{Aloisio05qt}. Secondly,
through the paper we identify the expectation value of the energy as
the temperature of black holes, which is a standard relation in
statistics for radiation particles. However, MDR may also change
this relation into $E\sim T(1+\delta lp^2T^2)$ with other correction
terms \cite{Alexander:2001ck,Ling06az}. Therefore, more exact
results of the Hawking temperature maybe have to take these
modifications into account. However, a delicate calculation shows
that this modification will not change the main conclusions we
present above at all.

\begin{acknowledgments}
We would like to thank all members, specially Xiang Li and Qingzhang
Wu, of the Center for Gravity and Relativistic Astrophysics at
Nanchang University for helpful discussions. This work is partly
supported by NSFC(Nos.10405027, 10663001), JiangXi SF(Nos. 0612036,
0612038), the key project of Chinese Ministry of Education (No.
208072) and Fok Ying Tung Eduaction Foundation(No.111008). We also
acknowledge the support by the Program for Innovative Research Team
of Nanchang University.
\end{acknowledgments}

\bibliography{apssamp}% Produces the bibliography via BibTeX.

\end{document}